# Superconductivity and Magnetism in $REFeAsO_{1-x}F_x$ (RE=Rare Earth Elements)

**K Kadowaki[1,2,3], T Goya[1], T Mochiji[1] and S V Chong[2]**

[1] Graduate School of Pure and Applied Sciences, University of Tsukuba, 1-1-1, Tennodai, Tsukuba, Ibaraki 305-8571, Japan

[2] Institute of Materials Science, University of Tsukuba, 1-1-1, Tennodai, Tsukuba, Ibaraki 305-8573, Japan

[3] CREST-JST, JSPS CTC-Program "NES" and WPI Center for Materials Nanoarchitechtonics (MANA), University of Tsukuba, 1-1-1, Tsukuba, Ibaraki 305-8573, Japan

E-mail: kadowaki@ims.tsukuba.ac.jp

**Abstract**. Fluoride-doped iron-based oxypnictides containing rare-earth gadolinium ($GdFeAsO_{0.8}F_{0.2}$) and co-doping with yttrium ($Gd_{0.8}Y_{0.2}FeAsO_{0.8}F_{0.2}$) have been prepared via conventional solid state reaction at ambient pressure. The non-yttrium substituted oxypnictide show superconducting transition as high as 43.9 K from temperature dependent resistance measurements with the Meissner effect observed at a lower temperature of 40.8 K from temperature dependent magnetization measurements. By replacing a small amount of gadolinium with yttrium $T_c$ was observed to be lowered by 10 K which might be caused by a change in the electronic or magnetic structure since the crystal structure was not altered.

## 1. Introduction

Recent discovery of Fe-based rare earth (RE) oxypnictide high temperature superconductors has attracted much attention, because their transition temperatures are surprisingly high, as high as 56 K at the present stage of research as reported in the systems with RE=Sm and Gd [1,2], and they contain strongly magnetic Fe element as an essential constituent for the electric conduction. In fact, the non F doped samples show strong magnetism, perhaps antiferromagnetism, which seems to disappear in the superconducting samples. This interplay of superconductivity and magnetism in these systems provides an interesting novel playground to study the mechanism of cuprate as well as non-cuprate high $T_c$ superconductivity. We have synthesized a series of samples with varying RE elements and here we report on the superconductivity in Gd and $Gd_{(1-x)}Y_x$ (Gd substituted with Y) electron-doped iron oxyarsenides.

## 2. Experimental

Polycrystalline $GdFeAsO_{0.8}F_{0.2}$ was prepared by reacting stoichiometric amount of GdAs, FeAs, Fe, $Gd_2O_3$, and $GdF_3$. GdAs was prepared by reacting gadolinium pieces and arsenic chips heated sequentially at 500 °C for 15 hours, 850 °C for 5 hours, and finally at 900 °C for 5 hours. FeAs was prepared by reacting pressed pellet of finely ground 1:1 molar ratio mixed powders of iron and arsenic at 700 °C for 10 hours. The starting materials were finely ground and thoroughly mixed before being pressed into pellets. The doping of Gd by a small amount of yttrium was carried out by adding $YF_3$ instead of $GdF_3$. The prepared pellets were placed in thoroughly dried quartz tubes and sealed under vacuum at low $10^{-2}$ Torr. The whole sample preparation was performed in pure argon gas filled glove-box. The pellets were than reacted sequentially first at 500 °C for 15 hours, 850 °C for 5 hours, 1000 °C for 24 hours before being transferred to a high temperature furnace and reacted at 1200 °C for 45 hours. The crystal structure of the resulting compound was determined by powder x-ray diffraction (XRD) using Cu Kα radiation at room temperature. Electrical resistivity was measured by a standard 4-probe technique down to 4 K. Magnetization measurements were carried out in a Quantum Design SQUID magnetometer down to 2 K.

## 3. Results and Discussion

Early study on Gd-oxypnictide has shown that when $Fe_2O_3$ was used as the oxygen source, the resulting electron doped compound was not superconducting although a drop in resistivity was observed at around 10 K [3]. Later on Chen *et al.* repeated the sample experiment but using $Gd_2O_3$ as the oxygen source which yielded the first Gd-based oxypnictide superconductor with a reported $T_c$ of around 36.6 K doped with 0.17 fluoride doping [4]. In this work temperature dependence of the mass susceptibility measurement (*M-T*) at 50 Oe applied field on $GdFeAsO_{0.8}F_{0.2}$ shows both magnetic screening and Meissner effect with an $T_c$ onset of around 40.8 K. Temperature dependent resistance measurement on the same sample shows an $T_c$ onset at as high as 43.9 K with $T_c$ at mid-point at around 42.0 K giving this the highest $T_c$ for Gd-based oxypnictide prepared at ambient pressure. There has since been a report on superconductivity in a similar sample $GdFeAsO_{0.8}F_{0.2}$ prepared under high pressure with $T_c$ onset at 51.2 K [5]. We have also confirmed the reagent dependent nature of this Gd-oxypnictide. When this compound was prepared with $Fe_2O_3$, the resulting *M-T* curves display only Curie-Weiss-like paramagnetic behavior down to low temperature as shown in the upper curves of Fig. 1. *M-T* measurements at different fields indicate the superconducting transition is still visible up to 1000 G, although considerably broaden, with an obvious magnetic ordering below $T_c$ at around 8 K. This is most likely due to the ordering of the rare-earth gadolinium magnetic moment which was also observed in CeFeAsO studied in our laboratory [6].The high temperature region of the *M-T* curves in Fig. 2 follows the Curie-Weiss-like behavior, $1/(\chi_T - \chi_0) = (T + \theta)/C$, with the fitted data giving Curie-Weiss temperatures $\theta < 0$ (ca. -22.6 K; inset) strongly supports a predominant antiferromagnetic ordering in this fluoride doped Gd-based oxypnictide superconductor.

Upon substituting a small portion of the Gd in $GdFeAsO_{0.8}F_{0.2}$ with Y, $T_c$ was observed to be lowered by 10 K. This is counter intuitive as one would expect an enhancement in $T_c$ similar to that observed in $La_{1-x}Y_xFeAsO$ superconductors [7]. There might be two explanations to this – (1) the structural alteration of adding Y might not be significant, hence the lowering of $T_c$ might be an electronic (or magnetic) effect; and (2) high pressure study on $CeFeAsO_{1-x}F_x$ revealed a negative pressure effect on this oxypnictide [8]. Comparing the XRD patterns of $GdFeAsO_{0.8}F_{0.2}$ and $Gd_{0.8}Y_{0.2}FeAsO_{0.8}F_{0.2}$ in Fig. 4, we observed no significant differences between them in which case the former explanation might be more appropriate. A more detail doping effect study by Y and subsequent high pressure study might be able to elucidate this point.

In summary, rare-earth iron oxypnictides containing Gd and Gd substituted with Y displayed superconductivity when they were electron-doped with fluoride. $T_c$ was lowered when Y was substituted into Gd, but without significantly altering the crystal structure, suggesting the change was due of electronic or magnetic effect(s).

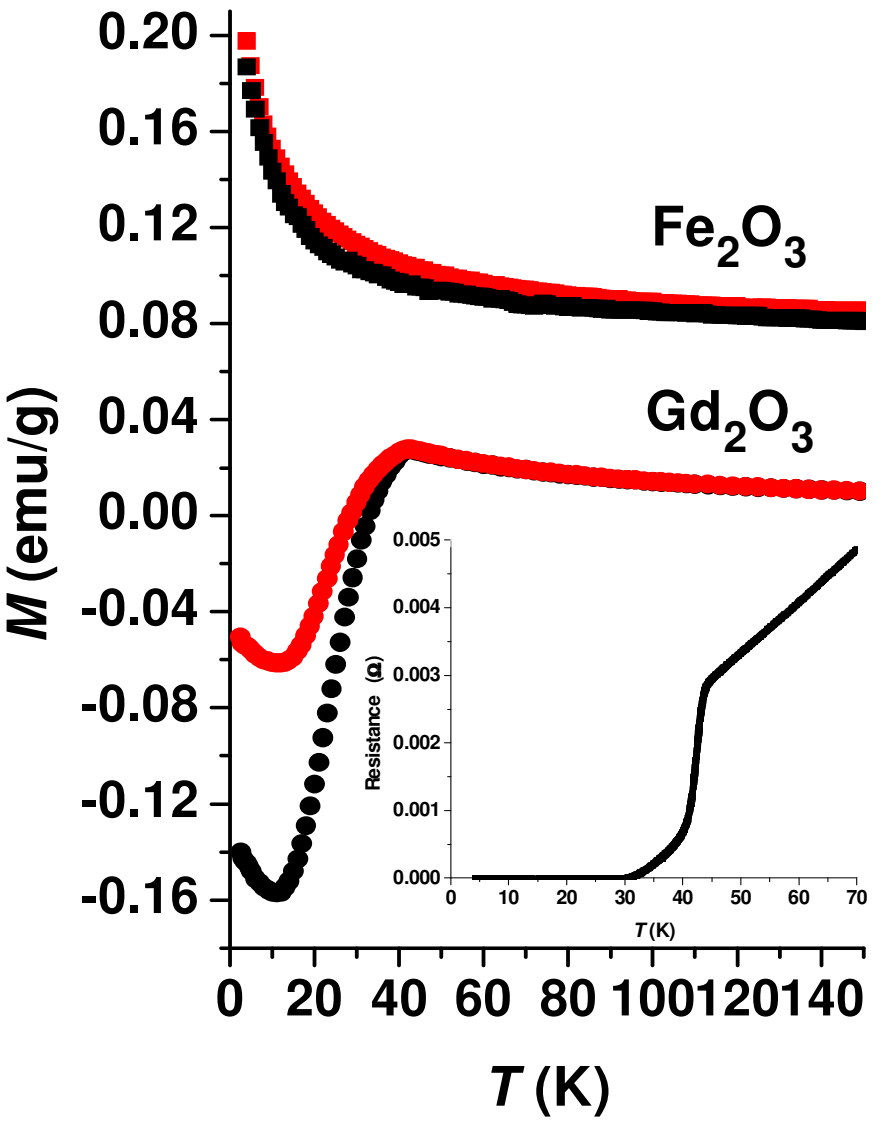

**Figure 1.** Temperature dependent mass susceptibility of the 0.2 F-doped sample prepared from different metal oxide sources.

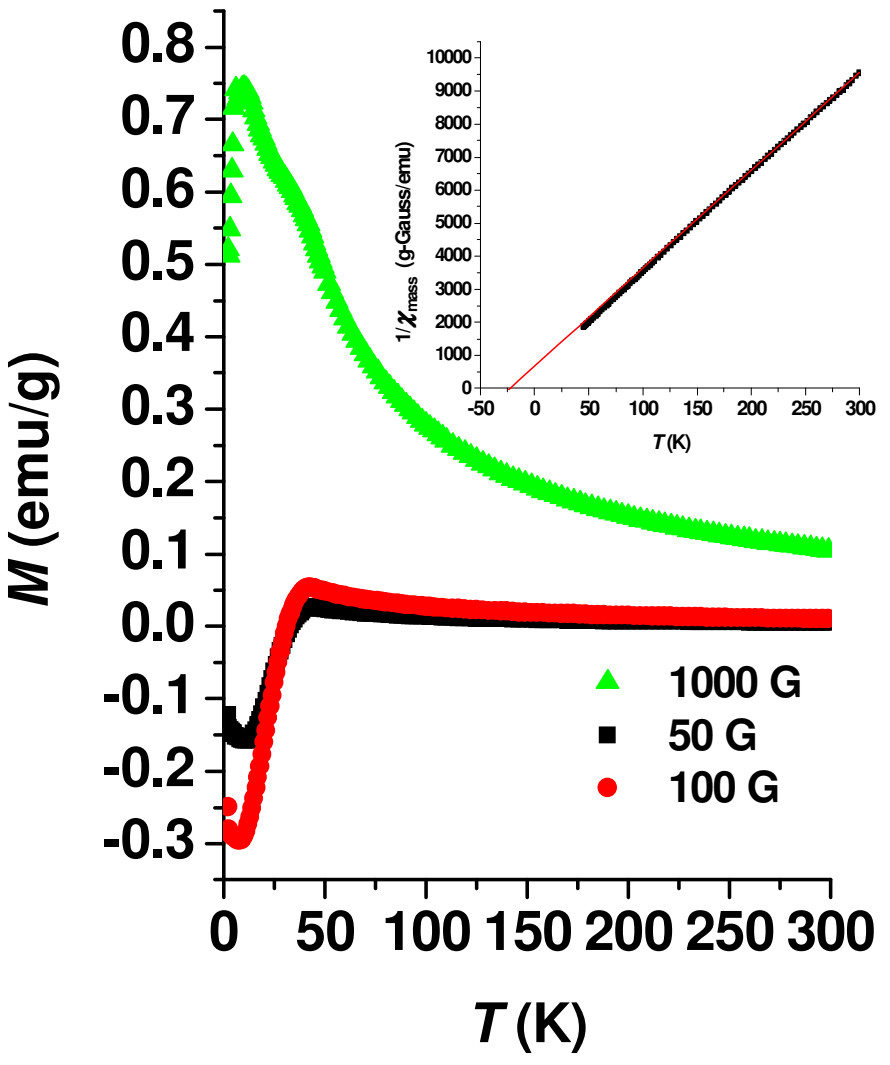

**Figure 2.** *M-T* of $GdFeAsO_{0.8}F_{0.2}$ measured at different field. The inset shows the Currie-Weiss fit on the 1000 G data above $T_c$.

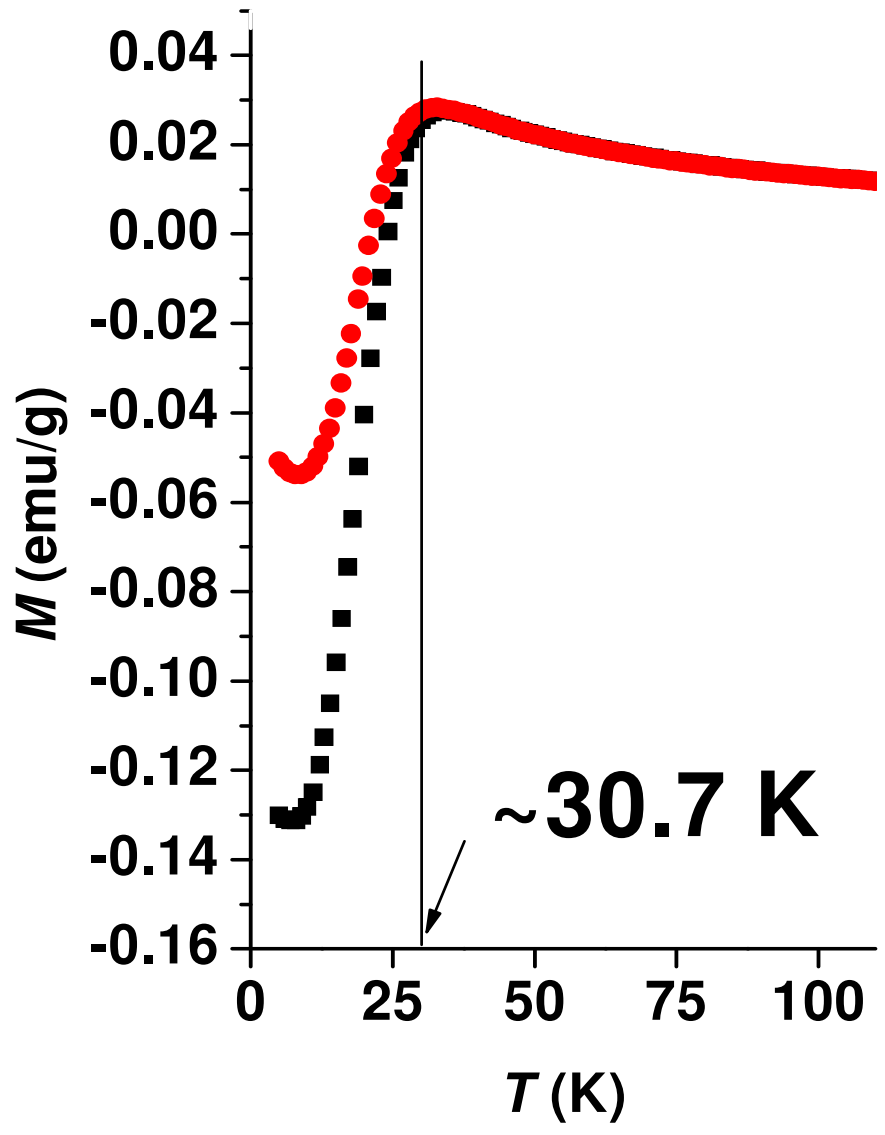

**Figure 3.** ZFC (lower) and FC (upper) measurements on $Gd_{0.8}Y_{0.2}FeAsO_{0.8}F_{0.2}$ superconductor.

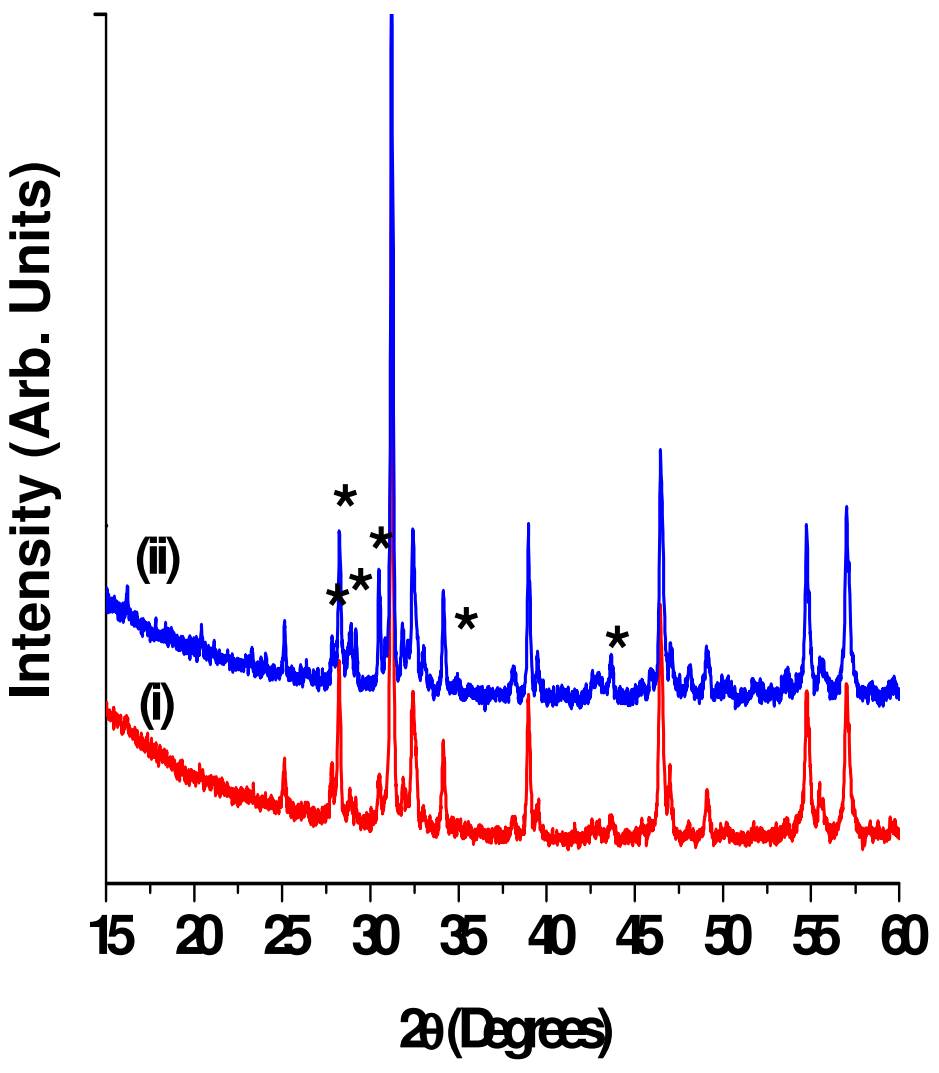

**Figure 4.** Powder XRD of (i) $GdFeAsO_{0.8}F_{0.2}$ and (ii) $Gd_{0.8}Y_{0.2}FeAsO_{0.8}F_{0.2}$. The asterisks (*) indicate the impurity phases in the compounds.